\documentclass[aps,pra,twocolumn,epsfig,showpacs]{revtex4}
\usepackage{graphicx}
\usepackage{amsmath}
\usepackage{latexsym}
\usepackage{amsfonts}
\usepackage{amssymb}
\usepackage{array}
\usepackage[dvips]{color}

\begin{document}

\title{
Entangled coherent states versus
entangled photon pairs for practical quantum information processing}

\author{
Kimin Park and Hyunseok Jeong} 
\affiliation{
Center for Macroscopic Quantum Control, 
Department of Physics and Astronomy,
Seoul National University, Seoul, 151-742, Korea}
\date{\today}

\begin{abstract}
We compare effects of decoherence and detection inefficiency on
entangled coherent states (ECSs) and entangled photon pairs (EPPs),
both of which are known to be particularly useful for quantum information processing (QIP).
When decoherence effects caused by photon losses are heavy,
the ECSs outperform the EPPs as quantum channels for teleportation
both in fidelities and in success probabilities.
On the other hand, when inefficient detectors are used, the teleportation scheme
using the ECSs suffers undetected errors that result in the degradation of fidelity,
while this is not the case for the teleportation scheme using the EPPs.  
Our study reveals the merits and demerits of the two
types of entangled states in realizing practical QIP under realistic conditions.  
\end{abstract}

\maketitle

\section{Introduction}
 
All-optical systems have been studied as a prominent candidate for physical
implementations of quantum information processing (QIP) \cite{Kok2007,Ralph2009}.
Quantum teleportation, which uses entangled quantum states as quantum channels,
plays a crucial role in optical quantum computation and
communication~\cite{KLMNature2001,GottesmanNature1999}.
One of the most difficult part in realizing quantum teleportation using optical systems is 
an efficient realization of the Bell-state measurement, 
in which four Bell states should be discriminated.
It was shown that the four Bell states cannot
be discriminated when only linear optical elements are used~\cite{Vaidman1999,LutkenhausPRA1999}, which 
makes it hard to achieve high success probability for quantum teleportation.
For example, in the teleportation scheme based on an entangled
photon pair (EPP)~\cite{BouwmeesterNature1997},
the success probability of the Bell measurement 
is bounded by $50\%$  when using only linear optical elements~\cite{ClsimigliaAPB2001}.
Even though universal gate operations can be realized based on
linear optics and photon detection~\cite{KLMNature2001},
this type of problem is one of the major hindrances 
to the implementation of deterministic gate operations as well as
scalable quantum computation.

An alternative qubit-based teleportation scheme was suggested \cite{vanEnkPRA2001,JeongPRA2001}
using  an entangled coherent state (ECS) as the quantum channel. 
In fact, the ECSs have been found to be useful not only for fundamental tests of quantum
theory~\cite{WilsonJMO2002}
but also for various applications in
QIP~\cite{vanEnkPRA2001,JeongPRA2001,Wang2001,JeongQIC2002,JeongPRA2002,
RalphPRA2003,BaAn03,LundPRL2008,BaAn09,Fiurasek10}.  
In this approach, a qubit is composed of two coherent states, $|\pm\alpha\rangle$,
where $\pm\alpha$ are the coherent amplitudes \cite{SchBook}.
It was explicitly pointed out in Refs.~\cite{JeongPRA2001,JeongQIC2002}
that all the four Bell states in the form
of ECSs can be well discriminated using only a beam splitter and 
two photon-number resolving detectors.
This has become a remarkable advantage in designing quantum computing schemes
using coherent-state qubits~\cite{JeongPRA2002,RalphPRA2003}
including deterministic gate operations with ECSs as off-line resources \cite{RalphPRA2003}.
Recently, it was shown that fault-tolerant quantum computing may be realized
with coherent-state qubits with amplitudes $\alpha > 1.2$~\cite{LundPRL2008}.

Implementations of high-fidelity EPPs and ECSs
in free-traveling fields
are challenging and crucial tasks for optical QIP. 
Recently, the realization of an electrically driven source of EPPs,
consisting of a quantum dot embedded in a semiconductor
light-emitting diode structure, has been reported
\cite{Shields2010}.
Even though the generation of high-fidelity ECSs is a demanding task,
remarkable experimental progress has recently been made
in generating single-mode superpositions of coherent
states~\cite{OurjoumtsevNature2007,SasakiPRL2008,Gerrits}, 
with which 
ECSs would easily be produced using an additional beam splitter.
Based on such progress,
several suggestions for the same purpose but higher fidelities and larger amplitudes
\cite{JJJ} have now become 
closer to the experimental realization.   
Efforts to generate arbitrary coherent-state qubits are also being made \cite{Neer2010}. 
Another difficult problem in the approach based on ECSs is that photon number resolving
detectors are required, while ongoing efforts are being made for the development of such
detectors~\cite{DaulerJMO2009,HadfieldNatPho2009}.

It is therefore important to compare the two optical QIP schemes,
one with single photon qubits and EPPs and the other with coherent-state qubits and ECSs, 
for efficient implementations of QIP in the long term. 
First, decoherence of quantum channels caused by photon losses
may be an obstacle against optical QIP.
This would be non-negligible particularly for long-distance quantum communication.
We therefore  study its effects on the two aforementioned teleportation schemes.
In general, when decoherence effect caused by photon losses is heavy
(or the decoherence time of the quantum channel is long),
the ECSs outperform the EPPs as quantum channels
both in teleportation fidelities and in success probabilities.
This tendency  becomes prominent
when the amplitude $\alpha$ is small: the ECSs outperform the EPPs regardless of
the decoherence time both in fidelities and in
success probabilities for $\alpha\lesssim 0.8$.

We also pay particular attention to the issue of detection inefficiency that
is a crucial detrimental factor 
in realizing practical QIP within all-optical systems. 
We point out that when inefficient detectors are used, the teleportation scheme
using ECSs suffers undetected errors that results in the degradation of fidelity.
This is not the case for the teleportation scheme using EPPs as
photon losses right before the detector errors are detected by the absence of 
the detection signals itself.  
We then present the results when both of the two factors, decoherence of the channel
and detection inefficiency, are applied.
Our results based on through quantitative analysis provide useful guidelines for
the choice of a scheme among well-known ones for practical QIP using optical systems.

\section{Decoherence of ECSs and EPPs}

In this section, we introduce the dynamics of ECSs and EPPs in a 
zero-temperature dissipative environment.
In this situation, photon losses occur that cause
the decrease of the average photon number and dephasing of the channels
at the same time. We discuss how the degrees of
entanglement for the ECSs and EPPs decrease by such decoherence effects.

\subsection{Solutions of master equation}

We are interested in ECSs in the form of \cite{Sanders92}
\begin{equation}
|\psi^\pm_\mathrm{ECS}\rangle=N_\alpha^\pm
\left(|\alpha\rangle_1|-\alpha\rangle_2\pm|-\alpha\rangle_1|\alpha\rangle_2\right)
\label{eq-ECS}
\end{equation}    
where $N_\alpha^\pm=1/\sqrt{2\pm2 e^{-4|\alpha|^2}}$ is the normalization factor.
The complex amplitude $\alpha$ is assumed to be real throughout the paper 
for simplicity without losing generality. 
We shall call
 $|\psi^+_\mathrm{ECS}\rangle$ ($|\psi^-_\mathrm{ECS}\rangle$) even (odd) ECS as 
 it contains only even (odd) numbers of photons. 
We also consider an EPP,
\begin{equation}
|\psi_\mathrm{EPP}\rangle=\frac{1}{\sqrt{2}}\Big(|H\rangle|V\rangle+|V\rangle|H\rangle\Big),
\label{EPPs}
\end{equation}
where $|H\rangle$ and $|V\rangle$ refer to horizontal and vertical polarization states,
respectively. 
The relative sign between the vector components 
of the EPP in Eq.~(\ref{EPPs}) was chosen to be $+1$ for simplicity:
this sign does not make any meaningful difference in our study and this is obviously different from the cases 
of the ECSs in (\ref{eq-ECS}) for which the signs in the middle play important roles. 
We also note that $|H\rangle$ is equivalent to 
$|1\rangle |0\rangle$ and $|V\rangle$ to $|0\rangle |1\rangle$
in terms of the dual-rail logic QIP.

The time evolution of density operator $\rho$ under
the Born-Markov approximation is given
by the master equation~\cite{PhoenixPRA1990}
\begin{equation}
\label{eq:me}
\frac{\partial \rho}{\partial \tau}=\hat{J}\rho+\hat{L}\rho,
\end{equation}
where $\tau$ is the interaction time, $\hat{J}\rho=\gamma\sum_ia_i \rho a_i^\dagger$,
$\hat{L}\rho=-\sum_i\frac{\gamma}{2}(a_i^\dagger a_i \rho + \rho a_i^\dagger a_i)$,
$\gamma$ is the decay constant, and $a_i$ is the annihilation operator for  mode $i$. The formal 
solution of Eq.~(\ref{eq:me}) is written as $\rho(\tau)={\rm exp}  [(\hat{J}+\hat{L})\tau]\rho(0)$, 
where $\rho(0)$ is the initial density operator. 
Assuming a zero-temperature bath, we obtain the density operator of the odd and even ECSs 
decohered in the vacuum environment as \cite{JeongPRA2001,vanEnkPRA2001}
\begin{align}
\label{eq:deco}
\rho^{\pm}_\mathrm{ECS}(\tau)=&(N_\alpha^\pm)^2\Big\{|t\alpha\rangle_1\langle 
t\alpha|\otimes|-t\alpha\rangle_2\langle -t\alpha|\\
+&|-t\alpha\rangle_1\langle -t\alpha|\otimes|t\alpha\rangle_2\langle t\alpha|\nonumber\\
\pm&e^{-4\alpha^2r^2}\big(|t\alpha\rangle_1\langle -t\alpha|
\otimes|-t\alpha\rangle_2\langle t\alpha|
+h.c.
)\Big\}  \nonumber
\end{align}
where $t = e^{-\gamma\tau /2}$
and superscript $+$ ($-$) corresponds to the even (odd) ECS.
We define the normalized time as $r = (1 - t^2 )^{1/2}$ for later use.
In what follows, we shall use only the {\it odd} ECSs,
which are maximally entangled in the $2\otimes2$ Hilbert space at time $\tau=0$,
as the quantum channels to teleport coherent-state qubits. As we shall
explain later, the odd ECS shows larger success probabilities of teleportation 
than the even ECS.
The density matrix $\rho^-_\mathrm{ECS}$ expressed in the orthogonal basis set 
$|\pm\rangle=N_\pm\big(|t\alpha\rangle\pm|-t\alpha\rangle\big)$ is given as
\begin{equation}
\label{eq:density matrix}
\mathbf{\rho^-_\mathrm{ECS}(\tau)}=\frac{1}{4(-1+e^{4\alpha^2})}\left(
\begin{array}{cccc}
A & 0 & 0 & D \\
0 & B & -B & 0 \\
0 & -B & B & 0 \\
D & 0 & 0 & C 
\end{array} \right),
\end{equation}
where 
\begin{align}
\label{eq:components}
A&=e^{-4(-1+r^2) \alpha^2}(-1+e^{4r^2\alpha^2})(1+e^{2(-1+r^2)\alpha^2})^2,\nonumber\\
B&=-1+e^{4\alpha^2}-e^{4r^2\alpha^2}+e^{-4(-1+r^2)\alpha^2},\nonumber\\
C&=e^{-4(-1+r^2) \alpha^2}(-1+e^{4r^2\alpha^2})(-1+e^{2(-1+r^2)\alpha^2})^2,\nonumber\\
D&=-1-e^{4\alpha^2}+e^{4r^2\alpha^2}+e^{-4(-1+r^2)\alpha^2}.
\end{align}

Using the same master equation, one can also find the density operator of the EPP for general $\tau$, 
initially given as
$\rho_\mathrm{EPP} (0) \equiv |\psi_\mathrm{EPP}\rangle\langle \psi_\mathrm{EPP}|$ at $\tau=0$, 
\begin{align}
\rho_\mathrm{EPP} (\tau)=&e^{-2\gamma\tau}\rho_\mathrm{EPP} 	(0)-2
\left(e^{-2\gamma\tau}-e^{-\gamma\tau}\right)\rho_1\nonumber\\
&+\left(e^{-2\gamma\tau}-2e^{-\gamma\tau}+1\right)\rho_v
\label{eq::decom}
\end{align}
where $\rho_1=\frac{1}{4}\sum_{i=1}^{4}|1\rangle_i \langle1|$ is a mixed single photon state density matrix, 
$|1\rangle_i\equiv|0\rangle\ldots|1\rangle_i \ldots|0\rangle$
 is a shorthand notation for a single photon occupying mode $i$ and the vacuum in all other modes, and 
$\rho_v$ represents the vacuum state for every mode. The density matrix can be represented in a basis set of 
$|H\rangle$, $|V\rangle$ and $|0\rangle$ similarly as before.
As one may expect, in a rough sense, the initial entangled two photon state decays 
to a mixed single photon state, and then eventually to the vacuum state.

\subsection{Degrees of entanglement}

As quantum teleportation utilizes entanglement as resource,
we first consider dynamics of entanglement for the ECSs and EPPs. 
Separability of a bipartite system is equivalent to the positivity 
of the partial transpose of the density matrix when the dimension
of the entire system does not exceed $6$ \cite{PeresPRL1996, HorodeckiPLA1996}.
We consider the ECSs in a 2 $\otimes$ 2 Hilbert space (using the dynamic qubit basis) as explained above even 
under the effect of photon losses. On the other hand, the EPPs evolve into 3 $\otimes$ 3 systems due to the 
addition of the vacuum element under photon loss effects. However, in our case of Eq. (\ref{eq::decom}), 
negativity of the total density operator equals the sum of negativities of all the decomposed components.  
This guarantees from the convexity of the negativity 
that this decomposition shows the smallest negativity  \cite{Vidal}.
It is known that the separability criterion is satisfied in such cases of the ``minimum decomposition'' 
\cite{SJLee}.

\begin{figure}
\includegraphics[width=0.5\textwidth]{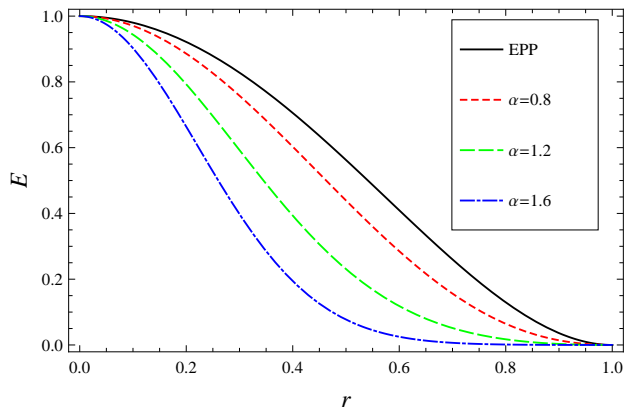}
\caption{ (Color online) Degrees of entanglement $E$ against the normalized time $r$.
The EPP shows larger entanglement than ECSs at any time regardless of $\alpha$.
}
\label{fig:negativity}
\end{figure}

Based on this, the measure of entanglement defined as $E=-2\sum_{i} \lambda_i^-$ can be used
\cite{LeeJMO2000}, where $\lambda_i^-$ are negative eigenvalues of the partial transpose of
the density operator.  Using Eqs.~(\ref{eq:density matrix}), (\ref{eq:components}) and
the abovementioned definition
of the entanglement measure, the degree of entanglement for an odd ECS is obtained as
\begin{align}
E_{\rm{ECS}}(\alpha,r)=-\frac{A+C-\sqrt{A^2+4B^2-2A C + C^2}}{4(-1+e^{4\alpha^2})},
\end{align}
and the degree of entanglement for the EPP is
\begin{equation}
\label{eq:EEPP}
E_{\rm EPP}(r)=(1-r^2)^2.
\end{equation} 
We have plotted the degrees of entanglement for the EPPs and ECSs
for several values of $\alpha$ in Fig.~\ref{fig:negativity}.
As it is already discussed \cite{LeeJMO2000,KimAndBuzekPRA1992}
the ECSs with large amplitudes decohere faster than those with small amplitudes.  
In the limit of $\alpha\rightarrow0 $, it is straightforward to show that
\begin{equation}
E_{\rm{ECS}}(\alpha,r)=-r^2+\sqrt{1-2r^2+2r^4}<E_{\rm{EPP}}(r)
\end{equation}
for $0<r<1$. Obviously, the EPP is always more
entangled than the ECS for any values of $\alpha$.

\section{Teleportation with ECS and EPP}

\begin{figure}
\includegraphics[width=0.5\textwidth]{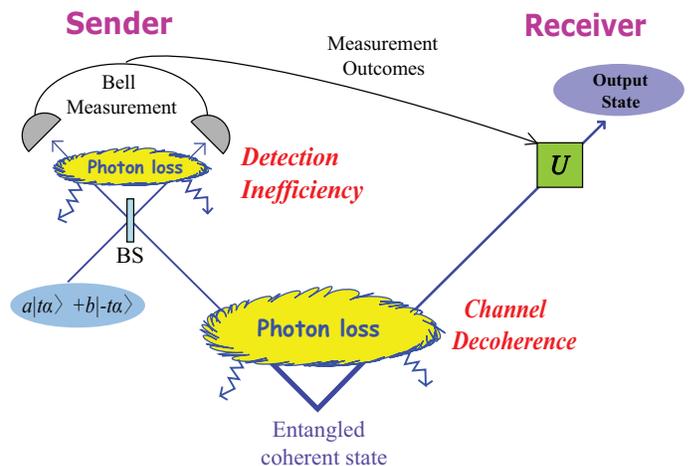}
\caption{Teleportation protocol using the ECS with two kinds of ``photon losses.''
Photon losses during the propagation of the quantum channel casuse the ``channel
decoherence'' while photon losses before ideal detectors
are introduced to model detection inefficiency.
BS represents a 50:50 beam splitter and $U$ the unitary operation required to 
restore the input state.
}
\label{fig:teleportation}
\end{figure}

It is obvious that with quantum channels decohered for non-zero decay time,
teleportation fidelities will degrade. This effect should not be neglected particularly
for long-distance quantum teleportation. Detection inefficiency may be an even more
crucial factor when considering practical quantum teleportation using optical systems.
It is often considered as photon losses in front of ideal detectors. 
We also note that dark count rates may be non-negligible for the cases
of highly efficient detectors such as photon number resolving detectors
necessary for the teleportation using the ECS. 
In this section, we thoroughly analyze the first two degrading factors
due to photon losses as depicted in Fig.~\ref{fig:teleportation}.

\subsection{Effects of channel decoherence}

The fidelity $F$ between input and output states for quantum teleportation is 
defined as $F=\langle\phi_\mathrm{in}|\rho_\mathrm{out}|\phi_\mathrm{in}\rangle$, where 
$|\phi_\mathrm{in}\rangle$ is the input state and $\rho_\mathrm{out}$ is the density
 operator of the output state.
For the case of an ECS, 
one can use $|t\alpha\rangle$ and $|-t\alpha\rangle$ as a dynamic qubit basis
in order to reflect amplitude losses as suggested in Ref.~\cite{JeongPRA2001}.
Then an unknown qubit reads 
\begin{equation}
|\phi_\mathrm{in}\rangle=a|t\alpha\rangle+b|-t\alpha\rangle
\label{eq:dq}
\end{equation}
where $a$ and $b$ are arbitrary complex numbers under the normalization condition.
The basis states $|t\alpha\rangle$ and $|-t\alpha\rangle$ are not orthogonal, but they 
approach such the limit for $t\alpha\gg1$. One can construct an 
orthogonal basis,
$|\pm\rangle=n_\pm 
(|t\alpha\rangle\pm|-t\alpha\rangle)$ with normalization factors $n_\pm$, 
 using their linear superpositions \cite{note1}.
In this way, one can consider the qubit (channel)
in a 2-dimensional ($2\otimes2$-dimensional) Hilbert space even under the decoherence effects. 
The input state can also be expressed as 
\begin{align}
|\phi_\mathrm{in} \rangle
=\cos({u/2})e^\frac{i v}{2}|+\rangle+\sin(u/2)e^{-\frac{i v}{2}}|-\rangle.
\end{align}
The coefficient $u$ and $v$ are
related to $a$ and $b$ as
 \begin{align}
& a=n_+\cos({u/2})e^\frac{i v}{2}+n_-\sin(u/2)e^{-\frac{i v}{2}},\nonumber \\
&b=n_+\cos({u/2})e^\frac{i v}{2}-n_-\sin(u/2)e^{-\frac{i v}{2}}.
\end{align}

The initial total state is then represented as
\begin{equation}
\rho^\mathrm{tot}=|\phi_\mathrm{in}\rangle_A\langle\phi_\mathrm{in}
|\otimes\{\rho_\mathrm{ECS}(\tau)\}_{BC},
\label{eq:totdensity}
\end{equation}
where $A$ and $B$ are modes for the sender while $C$ for the receiver. 
In order to discriminate between the Bell states,
a 50:50 beam splitter for modes $A$ and $B$ is used.
We here define the beam splitter operator as 
\begin{equation}
U_{i,j}(\theta)=e^{-\frac{\theta}{2}(a_i^\dagger a_j-a_i a_j^\dagger)}
\end{equation}
where $i$ and $j$ are two field  modes entering the beam splitter, and $\theta$ is related to the 
transmittivity $\zeta=\cos^2(\theta/2)$.
The action of the 50:50 beam splitter, $U_{A,B}(\pi/2)$, may be characterized as
$U_{A,B}(\pi/2)|\alpha\rangle_A|\beta\rangle_B=|(\alpha+\beta)/\sqrt{2} 
\rangle_A |(-\alpha+\beta)/\sqrt{2}\rangle_B$. 
The Bell states with coherent states in our context are 
\begin{eqnarray}
|\Phi^\pm\rangle=N_\pm (|t\alpha\rangle_1|t\alpha\rangle_2\pm|-t\alpha\rangle_1|-t\alpha\rangle_2), \\
|\Psi^\pm\rangle=N_\pm (|t\alpha\rangle_1|-t\alpha\rangle_2\pm|-t\alpha\rangle_1|t\alpha\rangle_2), 
\end{eqnarray}
where $N_\pm$ are normalization factors.
After the action of the beam splitter,
two photon number resolving detectors are required for modes $A$ and $B$
to complete the Bell-state measurement~\cite{JeongPRA2001}.
The projection operators $O_j$ for the outcomes $j$ representing the parity measurement
can be written as
\begin{eqnarray} 
\label{eq:O1}
O_1&=&\sum_{n=1}^\infty |2n\rangle_A\langle2n|\otimes|0\rangle_B\langle0|,\\
O_2&=&\sum_{n=1}^\infty |2n-1\rangle_A\langle2n-1|\otimes|0\rangle_B\langle0|,\\
O_3&=&\sum_{n=1}^\infty |0\rangle_A\langle0|\otimes|2n\rangle_B\langle2n|,\\
O_4&=&\sum_{n=1}^\infty |0\rangle_A\langle0|\otimes|2n-1\rangle_B\langle2n-1|,
\label{eq:O4}
\end{eqnarray}
where we refer to $\Phi^+$, $\Phi^-$, $\Psi^+$ and $\Psi^-$ as subscripts (or superscripts)
1, 2, 3 and 4 for simplicity.
In addition to the operators in Eqs.~(\ref{eq:O1}-\ref{eq:O4}), the error projection operator, 
$O_\mathrm{e}=|0\rangle_A\langle0|\otimes|0\rangle_B\langle0|$,
should also be considered because there is possibility for both the detectors not to register anything
even though such probability is very small when $\alpha$ is reasonably large. 

The unnormalized state after measurement outcome $j$ is obtained as
\begin{equation}
\rho^{j}=\mathrm{Tr}_{AB}[U_{A,B}(\pi/2)\rho_\mathrm{tot} U^\dagger_{A,B}(\pi/2)  O_j].
\end{equation}
Depending on the outcomes of the Bell-state measurement, different unitary rotations
on the coherent-state qubit for mode $C$ are required.
Applying an appropriate unitary operation $U_j$, 
the unnormalized output state is obtained as
$\rho^j_{\mathrm{out}}=U_j \rho^{j} U_j^\dagger$.
While no transformation or only a simple
phase shifter is required for the cases of $\Psi^-$ and $\Phi^-$,
the displacement operator is required for the other two cases that
degrades the fidelity when $\alpha$ is small.
We simply exclude such ``fidelity-degrading'' cases in this paper as the success probability
with an ECS is always higher than that with an EPP even {\it without} those cases.

\begin{figure}
\includegraphics[width=0.5\textwidth]{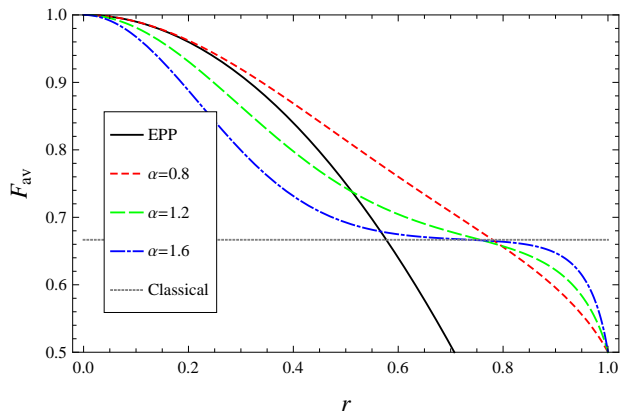}
\caption{
(Color online) The average teleportation fidelities, $F_\mathrm{av}$,
of the ECSs and the EPP as quantum channels against the normalized time $r$.
The dotted horizontal line indicates the maximum classical limit, 2/3,
which can be achieved by classical means.
}
\label{fig:teleportation fidelity}
\end{figure}

We find for the case of $\Psi^-$
\begin{align}
p_4 f_4&=\langle\phi_\mathrm{in}|\rho^4_{\mathrm{out}}|\phi_\mathrm{in}\rangle  \nonumber\\
&=(N^-_\alpha)^2e^{-2t^2\alpha^2}\sinh{(2t^2\alpha^2)}\nonumber\\
&\Big[|b|^2(a^*e^{-2t^2\alpha^2}+b^*)(ae^{-2t^2\alpha^2}+b)\nonumber\\
&+|a|^2(a^*+b^*e^{-2t^2\alpha^2})(a+be^{-2t^2\alpha^2})\nonumber\\
&+e^{-4\alpha^2(1-t^2)}a^*b(a^*e^{-2t^2\alpha^2}+b^*)(a+be^{-2t^2\alpha^2})\nonumber\\
&+e^{-4\alpha^2(1-t^2)}ab^*(a^*+b^*e^{-2t^2\alpha^2})(ae^{-2t^2\alpha^2}+b)\Big]\nonumber\\
&=p_2 f_2,
\end{align}
where $p_j=\mathrm{Tr}(\rho^j_\mathrm{out})$ is the probability of measuring a particular outcome $j$ 
and $f_j$ is the teleportation fidelity with that outcome. 
The success probability $p_4$ for $\Phi_-$ is obtained as
\begin{align}
p_4&=\mathrm{Tr}(\rho^4_\mathrm{out})\nonumber\\
&=
(N^-_\alpha)^2 e^{-2t^2\alpha^2} \sinh{(2t^2\alpha^2)}\nonumber\\
&~~~~\Big[|a|^2+|b|^2+e^{-4\alpha^2(1-t^2)}e^{-2t^2\alpha^2}(a^*b+ab^*)\Big]\nonumber\\
&=p_2.
\end{align}
The same calculations can be performed for 
the case of $\Psi_-$, which results in the same 
fidelity and the success probability.
The average teleportation fidelity over all unknown input states and the success probability are
\begin{equation}
\label{ecs-av}
F_\mathrm{av}=\frac{1}{4\pi} \int _0 ^\pi \sin u du \int _0 ^{2\pi} dv \frac{\sum_j p_j f_j}
{\sum_j p_j},
\end{equation}

\begin{equation}
\label{probability-av}
P=\frac{1}{4\pi} \int _0 ^\pi \sin u du \int _0 ^{2\pi} dv {\sum_j p_j} 
\end{equation}
where the summations run over only 2 and 4 since we discard all the other cases.
One can show  by performing the integration in (\ref{probability-av}) that the average success probability for 
the ECS
is $P_\mathrm{ECS}=1/2$, regardless of $\alpha$.
As we perform the integration in (\ref{ecs-av}), we obtain the expression
\begin{align}
\label{fidelity-analytic}
F_\mathrm{ECS}(\alpha,r)=&2n \frac{l-m}{c}\nonumber\\
&+2n \frac{d^2(l-m)+2c^2 m}{c^3}\frac{\mathrm{arctanh}~d/c-d/c}{(d/c)^3},
\end{align}
where now $l=3e^{8\alpha^2}-5e^{4\alpha^2(r^2+1)}+5e^{4\alpha^2(2+r^2 )}-3e^{4\alpha^2(1+2 r^2)}$, 
$m=(e^{4\alpha^2}+e^{4r^2\alpha^2})(e^{4\alpha^2}-e^{4\alpha^2(1+r^2))}$, $n=e^{-4\alpha^2(1+r^2)}/16$, 
$c=e^{4\alpha^2}-1$ and $d=-e^{2(1+r^2)\alpha^2}+e^{-2(-1+r^2)\alpha^2}$.
We have plotted 
the results in Fig.~\ref{fig:teleportation fidelity}. 

\begin{figure}
\includegraphics[width=0.5\textwidth]{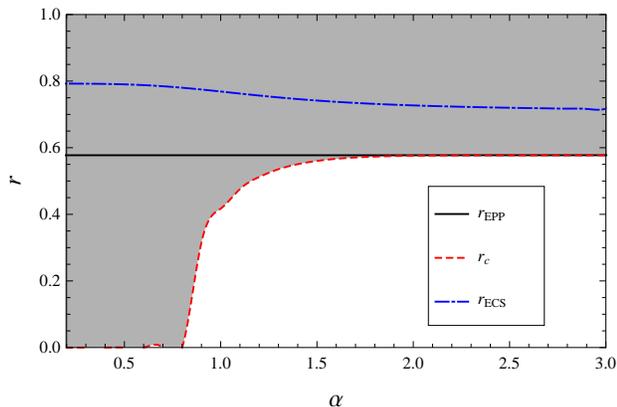}
\caption{
(Color online) The average fidelity using the EPP falls below the classical limit at $r_\mathrm{EPP}$ (solid 
line). The average fidelity using the ECS, $F_\mathrm{ECS}$, becomes larger than
that using the EPP, $F_\mathrm{EPP}$, at time $r_\mathrm{c}$ and falls below the classical limit at 
$r_\mathrm{ECS}$.
The grey shaded  area corresponds to $F_\mathrm{ECS}>F_\mathrm{EPP}$. }
\label{fig:domain}
\end{figure} 

The calculations are straightforward for the case of the EPP
because of the orthogonal nature of the qubit
and the channel.
In this case, only two ($|\Psi'^+\rangle$ and $|\Psi'^-\rangle$)
among the four Bell states,
$|\Phi'^\pm\rangle= (|H\rangle_1|H\rangle_2\pm|V\rangle_1|V\rangle_2)/\sqrt{2}$
and $|\Psi'^\pm\rangle= (|H\rangle_1|V\rangle_2\pm|V\rangle_1|H\rangle_2)/\sqrt{2}$,
can be identified using linear optics elements and photodetectors.
This means that the success probability cannot exceed $1/2$
\cite{LutkenhausPRA1999}.
The average fidelity and the success probability can easily be obtained in the same manner
explained above as
$F_\mathrm{EPP}(r)=1-r^2$ and 
$P_\mathrm{EPP}=(1-r^2)/2$, respectively.
Here, it is immediately clear that $P_\mathrm{ECS}=1/2>P_\mathrm{EPP}$:
the success probability using the ECS is higher than that using the EPP
regardless of $\alpha$.

In Fig.~\ref{fig:teleportation fidelity}, the average fidelities for the ECS and the EPP,
$F_\mathrm{ECS}$ and $F_\mathrm{EPP}$ respectively, are plotted and compared.
The classical limit denoted by the horizontal dotted line in the figure is $2/3$,
under which quantum channels become useless for teleportation of qubits.
We find that the teleportation 
fidelities using the ECSs stay above the classical limit longer than those with the EPP
regardless of the values of $\alpha$.
As shown in Fig.~\ref{fig:teleportation fidelity}, the EPP becomes useless for teleportation 
at time $r_\mathrm{EPP}=1/\sqrt{3}\approx 0.577 $
while the ECSs become useless at $r_\mathrm{ECS}$,
where $r_\mathrm{ECS}$ is determined between $0.7$ and $0.8$ depending on $\alpha$.
We have investigated the cases for large values of $\alpha$ up to 4,
and our numerical results 
lead us to conjecture that $r_\mathrm{ECS}$ converges to $\sim 0.7$
when $\alpha$ becomes large.
As shown in Fig.~\ref{fig:domain}, $F_\mathrm{ECS}$
remains lower than $F_\mathrm{EPP}$ until the decoherence time $r$ becomes $r_c$.
When the decoherence time reaches $r_c$, $F_\mathrm{ECS}$ exceeds  $F_\mathrm{EPP}$.
Of course, $F_\mathrm{ECS}$ eventually
falls below the classical limit at time $r_\mathrm{ECS}$ as we mentioned above.
Remarkably, 
$r_\mathrm{c}\approx0$ for  $\alpha \lesssim  0.8$, which means
that the ECSs outperform the EPP for these values of $\alpha$.

Even though the EPP is always more entangled than ECS (Fig.~\ref{fig:negativity}), 
it does not always mean higher teleportation fidelity (Fig.~\ref{fig:teleportation fidelity}). 
The reason for this can be understood as originated from the different dynamics
of the two channels under photon loss effects.
With the ECS channels, we have been able to minimize the degradation of the 
teleportation fidelity using the dynamic qubit basis \cite{JeongPRA2001}.
This is not possible with the EPP. Photon losses cause the EPP to have the
``vacuum'' elements both at the sender's mode and at receiver's. 
In other words, the decohered EPP gets out of the initial $2\otimes 2$ Hilbert space
composed of $|H\rangle$ and $|V\rangle$ and
this ``escape'' for the EPP is a major difference from the case of the ECS.
The vacuum portion at receiver's mode, $C$, results in a significant decrease of the teleportation
fidelity. (On the contrary, in the following subsection, it becomes clear that the vacuum elements
at sender's modes, $A$ and $B$, are noticed by a failure of the Bell-state measurement 
and such an error can be discarded so that the fidelity is not affected.)


We here comment on the difference between the previous result in Ref.~\cite{JeongPRA2001}
and ours in this paper.
In Ref.~\cite{JeongPRA2001}, the time $r$ at which the teleportation fidelity of the ECS falls below the 
classical limit was independent of $\alpha$. In that paper, the singlet fraction of the channel state was used 
to calculate the optimal teleportation fidelity by the method suggested in Ref.~\cite{HorodeckiPRA1999}. 
However, this method is not optimized for the ECS under our decoherence model based on photon losses: when 
$\rho^-_\mathrm{ECS}$ is partially traced over one of the modes,
the reduced density matrix is not proportional to the identity matrix, which is the condition
required to apply the singlet fraction method presented in Ref.~\cite{HorodeckiPRA1999}.

So far, we have not considered the even ECS. Because of the same reason as the case of the odd ECS, only 
$\psi^+$ and $\phi^+$ can be considered the successful Bell measurement results.
For the results with the even ECS, the teleportation fidelity becomes identical to the case of the odd ECS. 
However, the success probability is lower than that of the odd ECS according to our calculation for the same 
value of $\alpha$. The reason for this is 
as follows. We utilize the results of odd photon detection for the case of the odd ECS, while the results of 
the non-zero even photon detection, corresponding to $\psi^+$ and $\phi^+$, are used for the case of the even 
ECS.  The odd photon detection probability is the same to the even photon detection probability when taking 
the average over all input states. However, the even photon detection probability contains the ``all-zero" 
cases, which are eventually discarded, and this inconclusive failure  probability gets larger as the amplitude 
becomes smaller. Therefore, the even ECS channel results in lower success probability unless 
$\alpha\rightarrow\infty$.

\subsection{Effects of detection inefficiency}

We now consider the inefficiency of detectors that is one of the major obstacles to the realization
of quantum teleportation using optical systems.
An inefficient detector can be modeled by inserting a beam splitter of transmittivity $\eta$ in front of the 
perfect detector,
where the beam splitter operation mixing the light with fictitious vacuum mode can be denoted as $U^\eta 
_{i,j}\equiv U_{i,j}(\theta_\eta)$ where $\theta_\eta=2\cos^{-1}\sqrt{\eta}$, where $i$ and $j$ are indices 
for modes. 
In order to perform the Bell-state measurement, we first need to apply the 50:50 beam splitter to 
the total density operator $\rho^\mathrm{tot}$ in Eq.~(\ref{eq:totdensity}).
The beam splitter operations, $U^\eta$, for inefficient detectors are then applied to incorporate detection 
inefficiency.
The resultant density operator after tracing out the irrelevant vacuum modes ($v_1$ and $v_2$) is 
\begin{align}
(\rho^\eta)_\mathrm{ABC} &={\mathrm Tr}_{v_1,v_2}\Big [U^\eta_{A,v_1} U^\eta_{B,v_2}  
U_{A,B}(\pi/2)\{(\rho^\mathrm{tot})_{ABC}\nonumber\\
&\otimes (|0\rangle \langle0|)_{v_1}\otimes (|0\rangle \langle0|)_{v_2}\} U^\dagger_{A,B}(\pi/2) 
U^{\eta\dagger}_{B,v_2} U^{\eta\dagger}_{A,v_1}\Big ].
\end{align}
The unnormalized density matrix for measurement outcome $j$ is given as
$\rho^j_{\mathrm out}=U_j [Tr_{AB}(\rho^\eta O_j)] U^\dagger_j$.
Using Eqs.~(18) and (19), we find
\begin{align}
p_2f_2&=p_4f_4=\langle\psi_{in}|\rho^4_{\mathrm out}|\psi_{in}\rangle\nonumber\\
&=D (N_\alpha^-)^2 \Big[|L|^2+|M|^2+2e^{-4\alpha^2 r^2}C^2 Re(M^*L)\Big]
\end{align}
\begin{align}
\label{eq:probabilityefficiency}
p_2&=p_4=Tr(\rho_{\mathrm out}^4)\nonumber\\
     &=D (N_\alpha^-)^2\Big[|a|^2+|b|^2+2e^{-2\alpha^2(1+r^2)}C^2 Re(a^*b)\Big],
\end{align}
where $D=e^{-2\eta (1-r^2)\alpha^2}\sinh{(2\eta (1-r^2)\alpha^2)}$, $C=e^{-2(1-r^2)\alpha^2(1-\eta)}$, $M=a^* 
(a+b e^{-2(1-r^2)\alpha^2})$ and $L=b^* (a e^{-2(1-r^2)\alpha^2}+b)$,
and the average fidelity is obtained using Eq.~(\ref{ecs-av}) as
\begin{align}
F_\mathrm{ECS}(\eta,\alpha,r)=&2n \frac{l-m}{c}\nonumber\\
&+2n \frac{d^2(l-m)+2c^2 m}{c^3}\frac{\mathrm{arctanh}~d/c-d/c}{(d/c)^3},
\end{align}
where now $l=3S^{2(1+\eta)}-5S^{2(r^2+\eta)}+5S^{2(2+r^2 \eta)}-3S^{2(1+r^2(1+\eta))}$, 
$m=(S^{2}+S^{2r^2})(S^{2\eta}-S^{2(1+r^2\eta)})$, $n=S^{-2(1+r^2\eta)}/16$, $c=S^{2}-S^{-2(-1+r^2)(-1+\eta)}$, 
$d=-S^{(1+r^2)}+S^{-(-1+r^2)(-1+2\eta)}$,
and $S=e^{-2\alpha^2}$.

\begin{figure}
\includegraphics[width=0.5\textwidth]{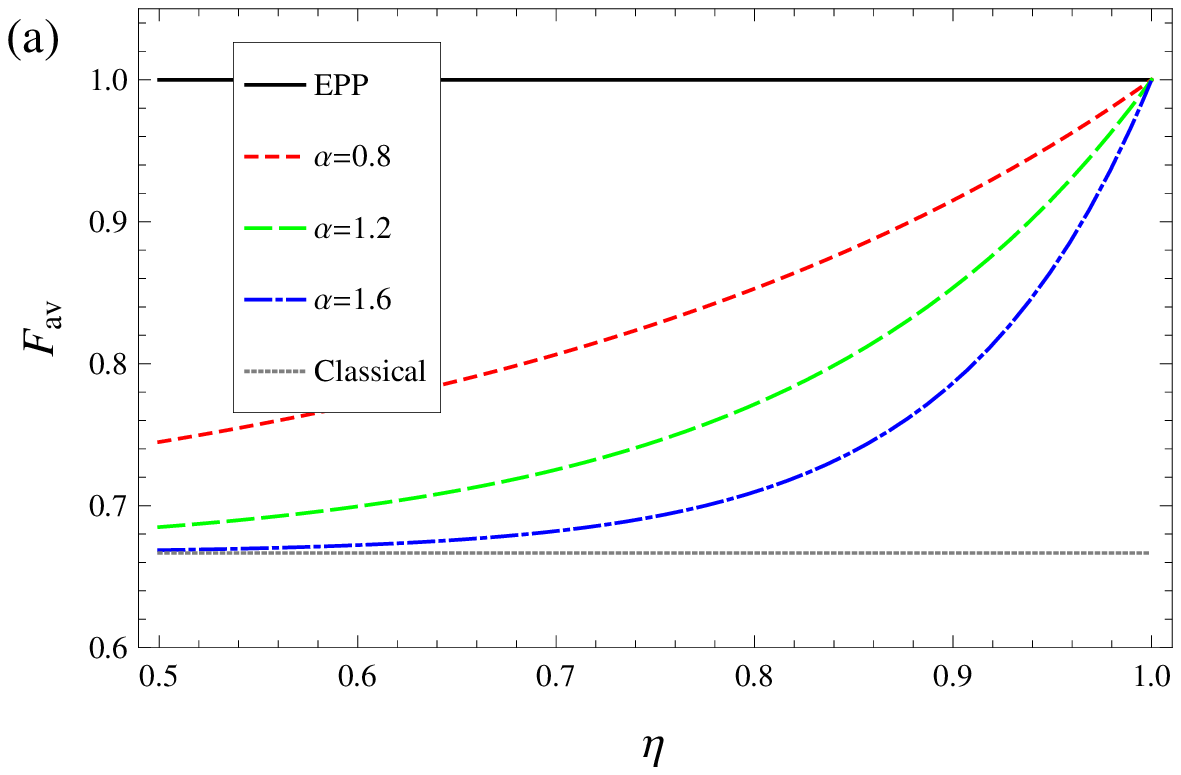}
\includegraphics[width=0.5\textwidth]{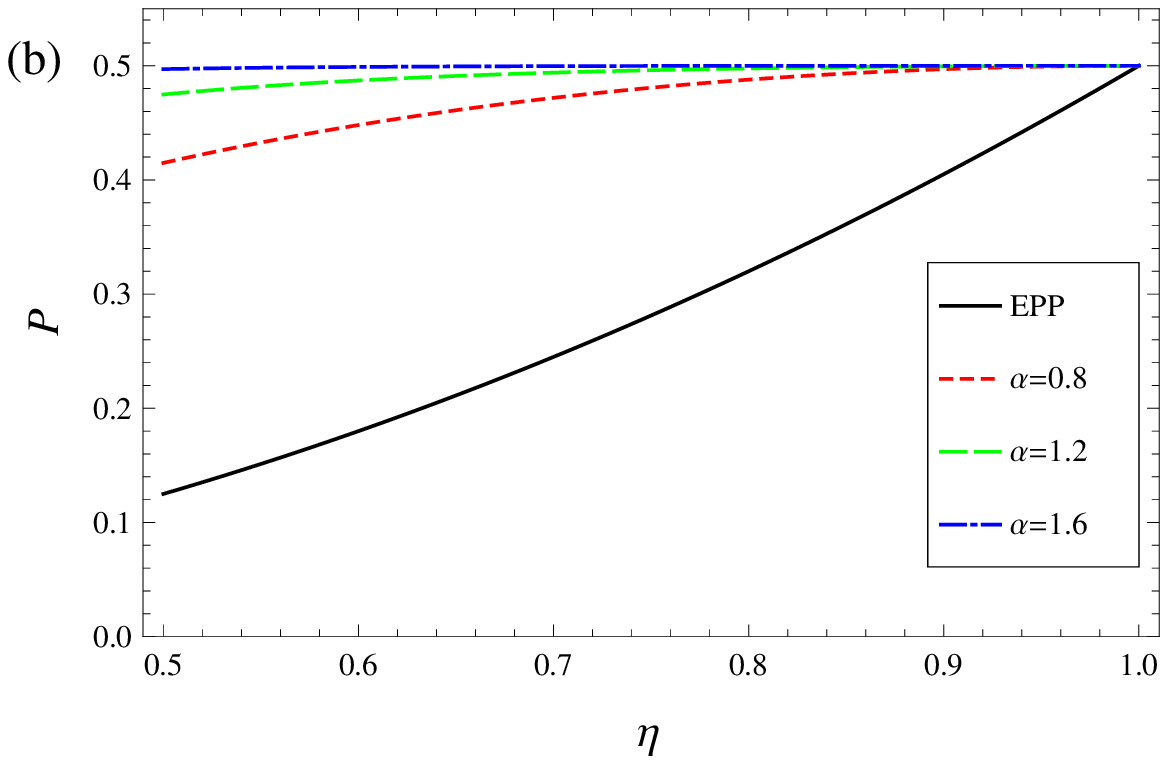}
\caption{
(Color online) (a) Teleportation fidelities using the ECS and the EPP as quantum channels 
in terms of the efficiency $\eta$ of detectors.
The ECS with large $\alpha$ shows smaller fidelity than that with small $\alpha$
while the fidelity using the EPP is not affected $\eta$.
(b) The success probabilites of teleportation using the ECS and EPP.
The success probability of the EPP decreases faster than that of the ECS by $\eta$. 
Decoherence of the channels is not considered to clearly see the effect of the 
detection inefficiency.}
\label{fig:Fvset}
\end{figure}

We first plot the teleportation fidelities for $r=0$ (i.e. without decoherence) in Fig.~\ref{fig:Fvset}(a).
It is clear that the ECSs with larger amplitudes are more sensitive to inefficiency 
of the detectors (i.e. decrease of $\eta$).
The reason for this is similar to the case of the channel decoherence.
The action of the beam splitter used for the Bell-state measurement may be described as
\begin{equation}
\begin{aligned}
&(a|\alpha\rangle+b|-\alpha\rangle)|\alpha\rangle\rightarrow
a|\sqrt{2}\alpha\rangle|0\rangle+b|0\rangle|\sqrt{2}\alpha\rangle,\\
&(a|\alpha\rangle+b|-\alpha\rangle)|-\alpha\rangle\rightarrow
a|0\rangle|-\sqrt{2}\alpha\rangle+b|-\sqrt{2}\alpha\rangle|0\rangle.
\end{aligned}
\end{equation}
It is then obvious that there are, for example,
``cross'' terms such as $|\pm\sqrt{2}\alpha\rangle\langle \mp\sqrt{2}\alpha |$ 
as well as the ``diagonal'' terms such $|\pm\sqrt{2}\alpha\rangle\langle
\pm\sqrt{2}\alpha|$ before the detection. 
Then, the cross terms described above in the density matrix 
are reduced as
$\propto e^{-4(1-\eta)\alpha^2}|\pm\sqrt{\eta}\sqrt{2}\alpha\rangle
\langle \mp\sqrt{\eta}\sqrt{2}\alpha 0|$ while 
the diagonal terms 
change simply to $\propto|\sqrt{\eta}\sqrt{2}\alpha\rangle\langle\sqrt{\eta}\sqrt{2}\alpha|$
due to photon losses modeled by beam splitters right in front of the ``perfect'' detectors. 
It is then straightforward to see that this reduction of the cross terms eventually causes
the teleported qubit to be mixed. 
Therefore, the inefficiency of the detectors (modeled by the additional beam splitters)
causes the teleported qubit to be ``more mixed'' when when the amplitude is larger.

On the contrary, the detection efficiency does not affect the teleportation fidelity using the EPP.
In this case, the number of photons that should be registered by the Bell measurement is
precisely defined as two. The Bell measurement succeeds only when two photons are registered by two of the 
four detectors used for the measurement \cite{LutkenhausPRA1999}.
If photon loss occurs due to the inefficiency of the detectors so that only one photon (or no photon at all) 
is detected, it will be immediately recognized by Alice as a failure. Alice can then simply filter out this 
kind of ``detected'' errors to prevent the decrease of the fidelity.

The success probability of teleportation using the ECS is obtained by 
Eq.~(\ref{probability-av}), $p_2$ and $p_4$ in Eq.~(\ref{eq:probabilityefficiency})  as:
\begin{align}
\label{eq:p-imperfect}
P_{\rm ECS}(\eta,\alpha,r)=&\frac{1}{4}S^{-2(-1+r^2)(-1+\eta)}(-1+S^{2(-1+r^2)\eta})\nonumber\\
&(-1+S^{2(1+(-1+r^2)(-1+\eta))})(-1+S^{2})^{-1}\nonumber\\
&(-1+S^{2(-1+r^2)})^{-1}.
\end{align}
The ECSs with small $\alpha$ show lower success probabilities than large $\alpha$
as seen in Fig~\ref{fig:Fvset}(b).
When $\alpha$ is small, even a small amount of photon losses
may significantly increase the possibility of $O_e$
(i.e., silence of both the detectors),
while this is not the case for large $\alpha$.
Therefore, the success probability using the ECSs with small $\alpha$ is more sensitive
to detection inefficiency, which is opposite to the case of the fidelity.
 
Of course, the ``filtering out'' of the detected errors for the case of the EPP
results in the more rapid decrease of the success probability.
The success probability using the EPP including the inefficient detector is similarly obtained
as for the case of the ECS as
\begin{equation}
P_\mathrm{EPP}(\eta,r)=\frac{1-r^2}{2}\eta^2
\end{equation}
and is plotted in Fig.~\ref{fig:Fvset}(b).
The additional factor $\eta^2$ when compared to the probability for the perfect detection case
means that each of the two photons in the Bell-measurement module is successfully 
detected with probability $\eta$. Here, we can easily check that the success probability
of the ECS is larger than the EPP regardless of $\alpha, r$ and $\eta$. Eq.~(\ref{eq:p-imperfect}) is reduced 
to $(2\eta + r^2 (-1+\eta) \eta -\eta^2)/2$ when $\alpha\rightarrow0$, and cannot be smaller than 
$P_\mathrm{EPP}(\eta,r)$ for any $\eta$ and $r$.

\subsection{Photon losses both in channels and at detectors}
\begin{figure}
\includegraphics[width=0.5\textwidth]{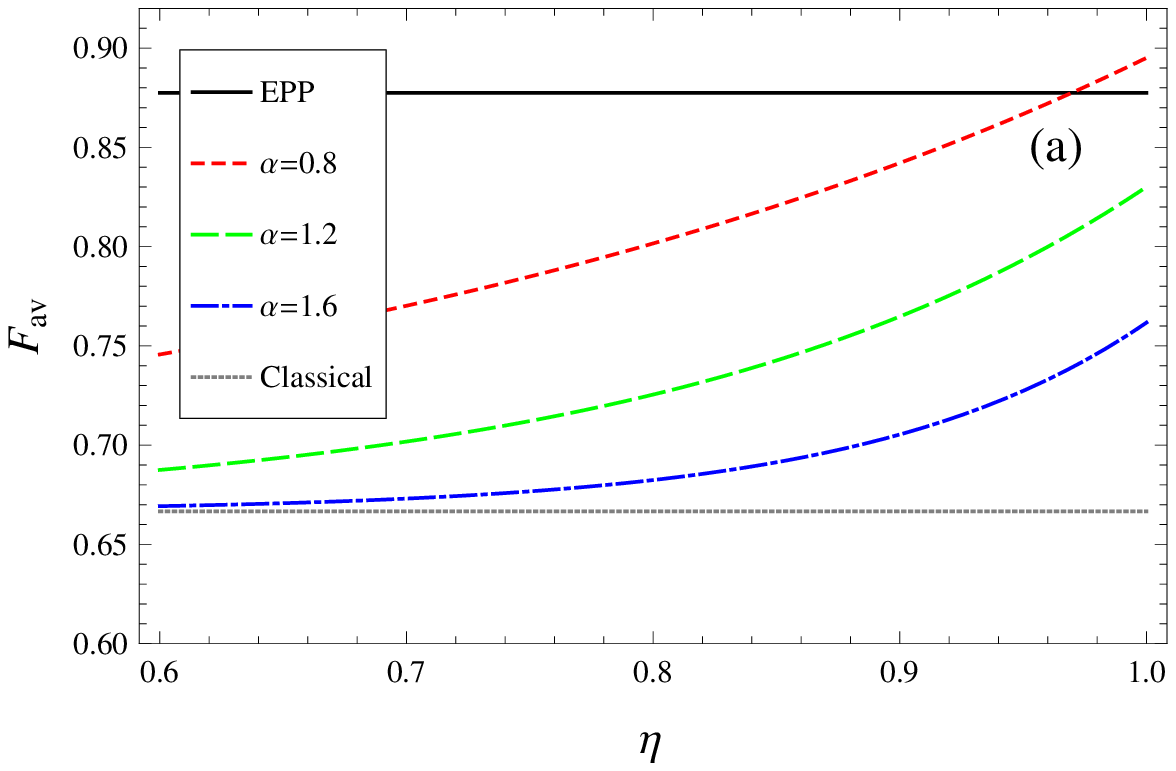}
\includegraphics[width=0.5\textwidth]{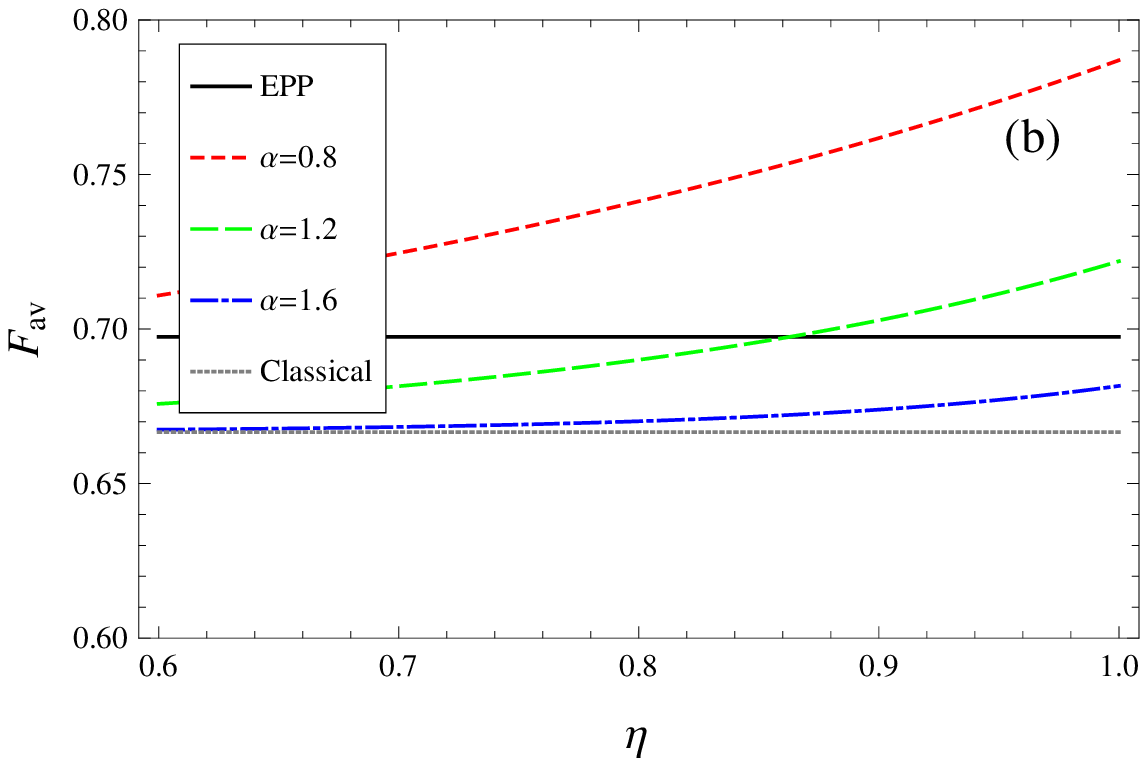}
\includegraphics[width=0.5\textwidth]{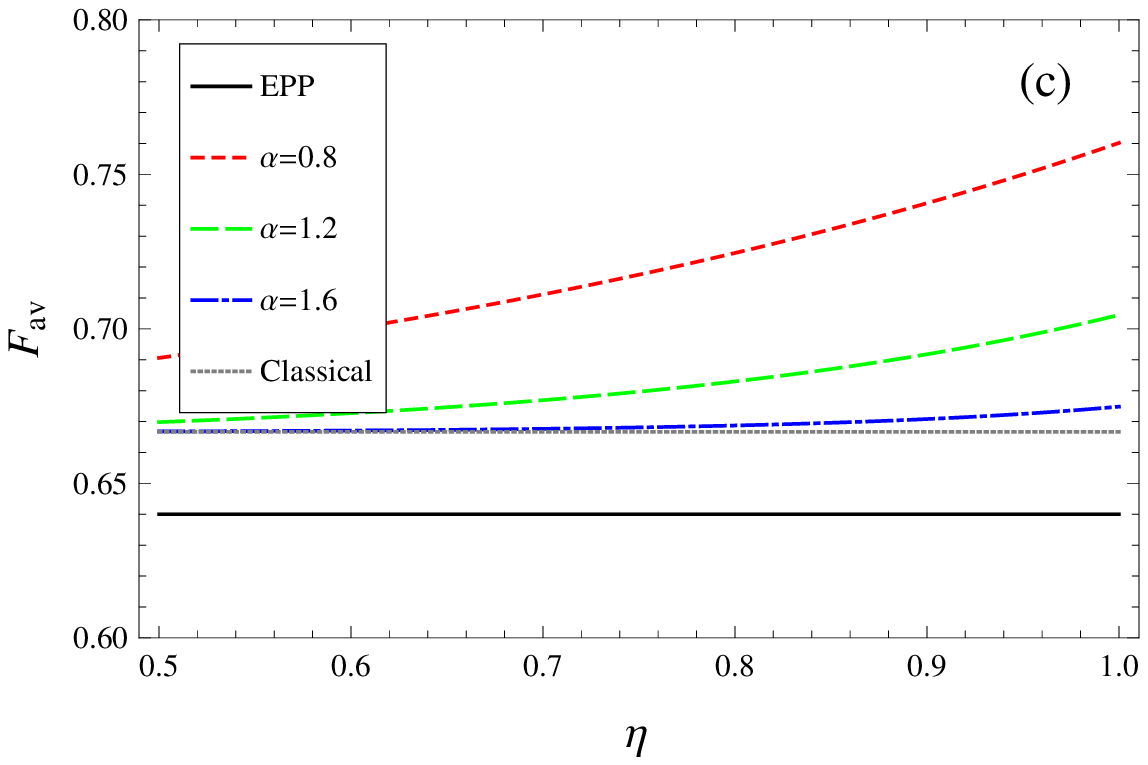}
\caption{
(Color online) Teleportation fidelities against detection efficiency $\eta$ at 
decoherence time (a) $r=0.35$, (b) $r=0.55$ and (c) $r=0.6$ for several values of $\alpha$.
As $r$ becomes larger, the fidelity with the EPP drops more rapidly than
the fidelities with the ECS.}
\label{fig:differentr}
\end{figure} 
So far, we have separately considered two different kinds of photon losses, the losses in the channel 
(referred to as channel decoherence) and the losses at the detectors (detection inefficiency) used for the 
Bell-state measurements. In realistic situations, both kinds of losses exist, and it is meaningful to know how 
the fidelities change under the combination of these effects.

If the ECS shows larger fidelity than the EPP with the perfect detector,
it is expected that this is true with imperfect detectors for some moderate values of $\eta$. 
As shown in several examples in Fig.~\ref{fig:differentr}, the ECSs begin to show larger fidelities
even with inefficient detectors as the decoherence time gets larger. 
As noted in the previous section, only the channel decoherence degrades the teleportation fidelity with the 
EPP, while the teleportation fidelity with the ECS is affected by both the channel decoherence and the 
detection inefficiency. When the decoherence effect is as dominant as $r>0.577$,
the teleportation fidelity with EPP becomes lower than the classical limit, $2/3$,
and the teleportation fidelities with the ECSs are always higher regardless of any other conditions.

\section{Conclusions}

In this paper, our attempt was to compare ECSs and EPPs as resources
for QIP under realistic conditions. 
We have considered decoherence caused by photon losses in ECSs and EPPs 
as quantum channels for teleportation.
We have pointed out that
entanglement of the EPPs is always larger than that of the ECSs in a dissipative environment.
On the other hand, the ECSs outperform the EPPs for the standard teleportation protocol
in fidelities for $\alpha\lesssim 0.8$.
Furthermore, the success probabilities for teleportation using the ECSs
are always higher than those using the EPPs.
However, as $\alpha$ gets larger, the range for which the EPPs show higher
fidelities appears.

In general, teleportation fidelity using the ECSs remains over the classical limit
longer than that of the EPPs. 
In other words, even when the EPPs become useless for teleportation
due to significant decoherence effects, the ECSs can still be useful for the same purpose.
Based on our numerical results we would conjecture that the ECSs are useful for teleportation
until the normalized time becomes $r\approx 0.7$ regardless of $\alpha$
while the EPPs become useless when $r\approx 0.577$.
However, when $\alpha$ is too large as, e.g., $\alpha>1.6$, this fidelity 
merit of the ECSs is too tiny so as to make the teleportation process useless.
We have thus pointed out that the degrees of decoherence in the quantum channels 
are a crucial factor to decide whether the ECSs or the EPPs should be used for efficient QIP.
On the other hand, it should be noted that the requirement for fault tolerant
quantum computing using coherent-state qubits is very demanding \cite{LundPRL2008}.


We also pay special attention to detection inefficiency that is a crucial
detrimental factor in realizing practical QIP using all-optical systems. 
We point out that when inefficient detectors are used for Bell-state measurements,
the teleportation scheme
using the ECSs suffers undetected errors that result in the degradation of fidelity.
This is not the case for the teleportation scheme using the EPPs as
photon losses right before the detector are noticed by the absence of 
the detection signals itself. 
Finally, we have presented analytical results and examples 
when both the channel decoherence and detection inefficiency are considered.
Our results based on a through quantitative analysis
reveal the merits and demerits of the two types of entangled states
in realizing practical QIP under realistic conditions, and
provide useful guidelines for the choice among the
well-known QIP schemes based on optical systems.

\acknowledgments

The authors thank Chang-Woo Lee for useful discussions.
This CRI work was supported by the National Research Foundation of Korea(NRF)
grant funded by the Korea government(MEST) (No. 3348-20100018).


\begin{thebibliography}{9}
\bibitem{note1} A direct comparison between arbitrary polarization qubits, $\mu|H\rangle+\nu|V\rangle$,
and coherent-state qubits, $\mu|\alpha\rangle+\nu|-\alpha\rangle$, is not straightforward when $\alpha$ is 
small due to the nonzero overlap between the two component coherent states. It should be noted that we use the 
orthogonal basis, $|+\rangle$ and $|-\rangle$, for coherent-state qubits in order to calculate the average 
fidelity for a ``fair comparison'' with the polarization-qubit case. In this sense, the input coherent-state 
qubit in comparison to the polarization qubit, $\mu|H\rangle+\nu|V\rangle$, should be considered to be 
$\mu|+\rangle+\nu|-\rangle$, rather than $\mu|\alpha\rangle+\nu|-\alpha\rangle$, when the overlap 
$|\langle\alpha|-\alpha\rangle|^2$ is non-negligible.
\end{thebibliography}
\end{document}